# Paraffin Wax Crystal Coarsening: Effects of Strains and Wax Crystal Shape

*Sasha Pechook,[†,‡] Alex Katsman,*[†] and Boaz Pokroy*[†,‡]*

[†]Department of Materials Science & Engineering, Technion  Israel Institute of Technology, 32000 Haifa, Israel

[‡]The Russell Berrie Nanotechnology Institute, Technion  Israel Institute of Technology, 32000 Haifa, Israel

**ABSTRACT**: We developed a model of paraffin wax crystal coarsening that well describes our experimental results and allows behavior of the paraffin films to be predicted on the basis of the extracted kinetic parameters. Wax crystalline films were evaporated on different substrates (silicon wafer, glass slide, thin layer of gold on silicon), thermally treated at different temperatures (4−60 ºC), and investigated by powder X-ray diffraction, high-resolution scanning electron microscopy, and optical confocal imaging of the surfaces. Preferred (110) crystal orientation of all deposited wax films, independently of substrate type, was observed from the start and increased during heat treatment. The change in preferred orientation was accompanied by changes in crystal morphology and shape, resulting in surface nano-roughening. We modeled the process as the coarsening of oriented $C_{36}H_{74}$ crystal islands driven by the decrease in total surface energy. Coarsening kinetics was controlled by diffusion of single molecular chains along the substrate. Evolution of nano-roughness during annealing time was well described by a surface coarsening law, $H_r \sim t^{1/4}$. Two additional factors influenced the evolution rate: strains accumulated in wax crystals during deposition, and divergence of initial crystal shape from the shape at equilibrium. Both factors lowered activation energy and effectively shortened coarsening time.

## INTRODUCTION

The n-alkanes are simple organic molecules, whose presence in various ordered molecular assemblies such as Langmuir-Blodgett films[1] and self-assembled monolayers[2] makes them useful in case studies of the orientation, assembly and morphology of complex organic films. In addition, n-alkanes have attracted research interest as materials for technological applications, as they can serve as lubricants, protective layers and surface modifiers controlling surface wettability and chemical properties. A tetratetracontane ($C_{44}H_{90}$) passivation layer, for example, was recently shown to increase charge carrier mobility in a field-effect transistor by an order of magnitude.[3,4]

Thin polyethylene films with improved mechanical properties were prepared as early as 1970.[5] Vapor deposition has since been shown to be a useful method for the formation of organic thin films with desirable molecular orientation, achievable through the control of deposition rate, substrate temperature, substrate surface energy and morphology.[6-8] Relatively high deposition rates and low substrate temperatures are commonly employed for thermal evaporation. These conditions lead to kinetically controlled growth, a non-equilibrium process that often terminates in structures considerably distorted from the bulk structures.[9]

Generally speaking, the formation of a thermally evaporated thin film can be divided into three steps: evaporation (from source material), adsorption to the substrate, and nucleation on the substrate. Adsorbed molecules can move along the substrate with alternating molecular orientations, in addition to undergoing re-desorption and diffusion processes that will determine the final molecular orientation and structure of the thin film (Figure 7a).[7] These processes have been extensively studied in thin films deposited via ultra-high vacuum growth by organic molecular beam deposition,[9] where many of the processes also occur in thermally evaporated films (moderate vacuum growth). In both cases, the growth of organic thin films is often a kinetic process that results in non-equilibrium structures leading to post-growth re-organization.[10] Theoretical studies have been restricted mainly to the description and prediction of growth modes and film morphology, particularly the scaling theory.[11,12] Nevertheless, both theoretical and experimental research are focused on growth stages and the resulting orientation and morphology.[6,13-17] In a recent study, Pechook et al. observed post- evaporation growth, from the nano to the micro scale at room temperature, of hexatriacontane ($C_{36}H_{74}$) wax crystals for wetting applications.[18] Up to now this phenomenon has not been further addressed. Here we present a thorough study of wax crystal growth and propose a theoretical model describing the suggested growth/coarsening.

## EXPERIMENTAL SECTION

### 1. *Sample preparation*

We prepared samples by thermal deposition of hexatriacontane wax, $C_{36}H_{74}$ (98%, Sigma-Aldrich (France)) on Si substrates, using a Bio-Rad Polaron Division Coating System. Deposition was carried out in a vacuum chamber at $10^{-5}$ mbar. Samples were positioned on a holder 10–12 cm above a crucible loaded with 10 mg of hexatriacontane wax. Evaporation occurred at ~120 ºC upon application of pulses of an electrical current. Deposition was conducted at ~20Å/s. The system was rapidly heated to ~120˚C by manual adjustment of the power. The evaporated specimens were transferred to 25 ºC (room temperature, RT). Thermal treatment procedures were conducted at 25 ºC, 32 ºC, 40 ºC, 50 ºC and 60 ºC for durations of 1, 5,10, and 50 h (MRC-1410DIG oven, in air).

### 2. *Characterization*

Wax powder and crystalline thin films were characterized structurally and micro-structurally by X-ray diffraction with a Cu-anode sealed tube (Rigaku, SmartLab X-Ray Diffractometer). Surface imaging was performed by HRSEM (Zeiss Ultra Plus). Micrographs were obtained at different times after deposition of the RT and heated samples. Roughness was assessed by confocal microscopy (Leica DCM 3D).

## EXPERIMENTAL RESULTS

Samples were prepared at a deposition rate of 20Å/s. Using high-resolution scanning electron microscopy (HRSEM) and optical confocal imaging of the surfaces, we investigated the morphology of wax crystalline films immediately after their deposition and after different thermal treatments (Figure 1). Examination of as-deposited films by X-ray diffraction (XRD) revealed the tensile stress accumulated in the layer after evaporation (Figure 2). The diffraction peak of the film is seen to have shifted to a higher diffraction angle, which corresponds to smaller d-spacing in the diffraction direction and larger d-spacing in the lateral direction (Poisson's ratio). Strains from the shift in the diffraction peak positions were determined as follows:

$$\varepsilon = \frac{d_m - d_0}{d_0} \qquad (1)$$

where $d_m$ and $d_0$ are the measured and the unstrained d-spacing respectively. Initial strains ε ≈ 1% decreased to zero during annealing (at 25 ºC for 50 h, or at 40 ºC for 10 h). A rapid roughening process was observed during annealing at temperatures above room temperature (25−50 ºC). Wax crystal shape changed during annealing, tending to become longer and narrower than the initial shape.

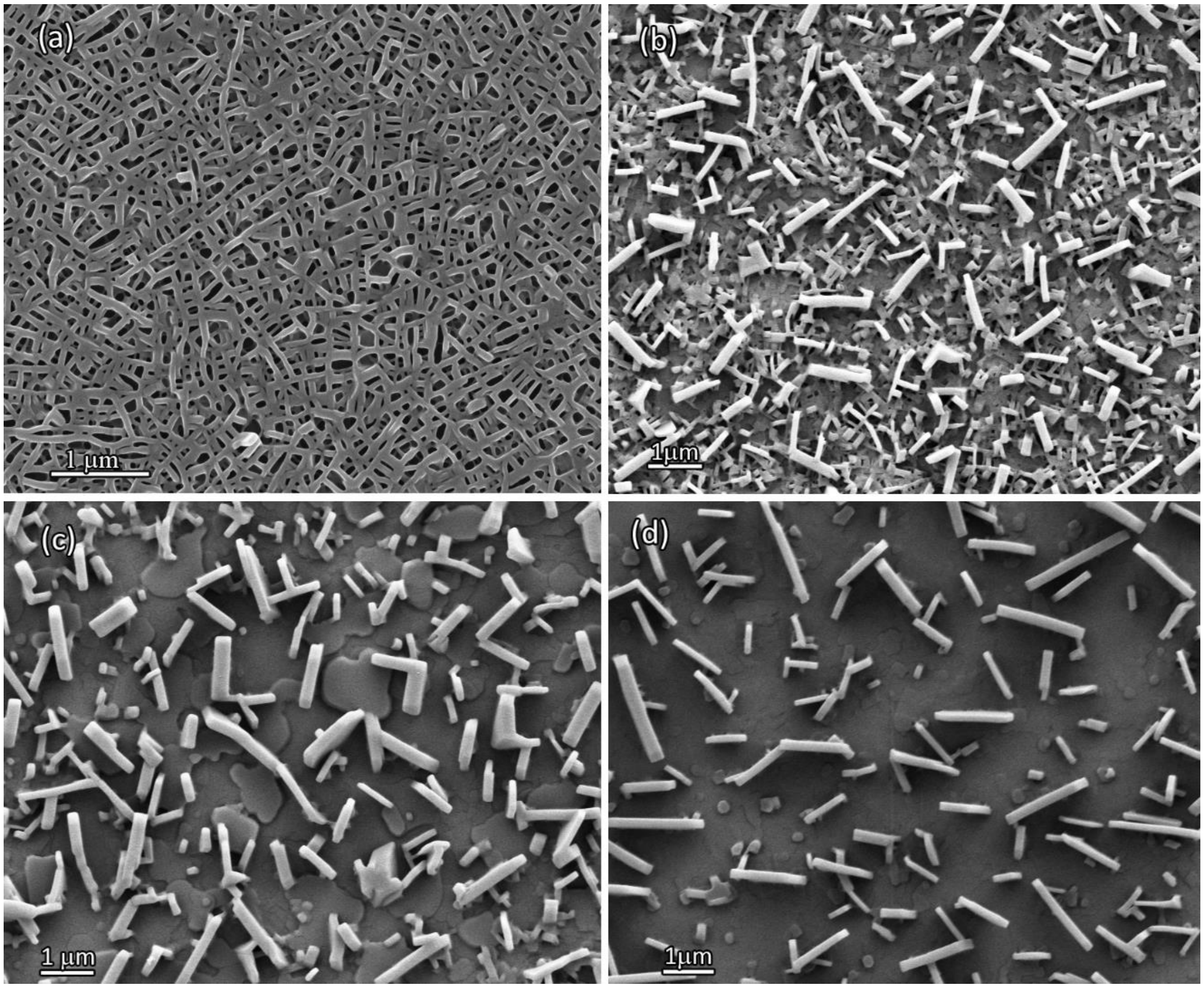

**Figure 1.** HRSEM images of thermally evaporated $C_{36}H_{74}$ wax on Si substrates at a deposition rate of 20 Å/s, (a) as a deposited layer, (b) after thermal treatment for 5 h, (c) after thermal treatment for 10 h, and (d) after thermal treatment for 50 h. Samples were thermally treated in air at 40 ºC.

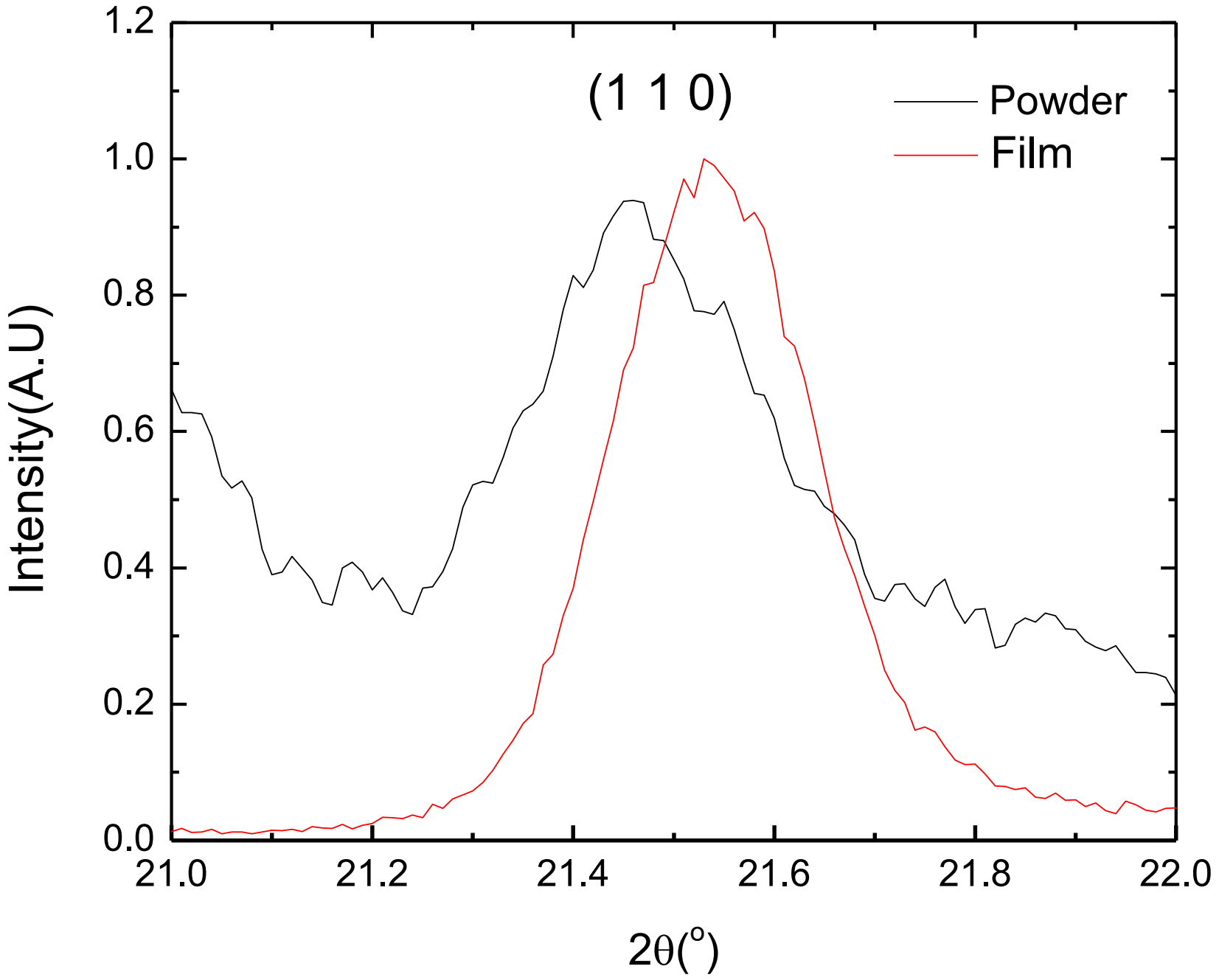

**Figure 2.** Comparison of the (110) diffraction peak with that of a powder sample in XRD spectra of thermally evaporated $C_{36}H_{74}$ on Si substrates.

Comparison of Figure 1a with Figure 1c, showing heat treatment-induced evolution of crystal shape, led us to view the growth mechanism at this stage in terms of crystal particle growth occurring at the expense of smaller particles via a coarsening process.

To better understand this process in the case of the wax crystals, we developed a theoretical model. Based on this model, we further analyzed our experimental results to determine the kinetic and thermodynamic parameters of the system.

## THEORETICAL MODEL

The $C_{36}H_{74}$ clusters that form islands on the deposited wax film are responsible for the measured initial nano-roughness, and their growth and coarsening correspond to evolution of wax film nano-roughness during thermal treatment. We consider the wax layer immediately below the islands as a substrate and a source of $C_{36}H_{74}$ molecules for growth of the islands, which then are considered as crystal wax particles. Let's assume that a wax crystal particle is a bar with sizes $W$, $H$ and $L$, where $W \sim H$, $L >> W$ and the size ratios $H/W = \kappa(\bar{\gamma}_s/\gamma_2)$ and $H/L = p(\bar{\gamma}_s/\gamma_1)$, where $\bar{\gamma}_s = (\gamma_{s0}+\gamma_s)/2$, $\gamma_s$ is the bar/substrate surface energy, $\gamma_{s0}$ is the energy of the external surface parallel to the substrate, $\gamma_1$ and $\gamma_2$ are corresponding interface energies of the bar, and κ and $p$ are the numerical coefficients (Figure 3). The values κ = $p$ = 1 correspond to the equilibrium shape of the bar, where the minimum surface energy determined by Wulff construction for the bar's volume $V_0 = HWL$.

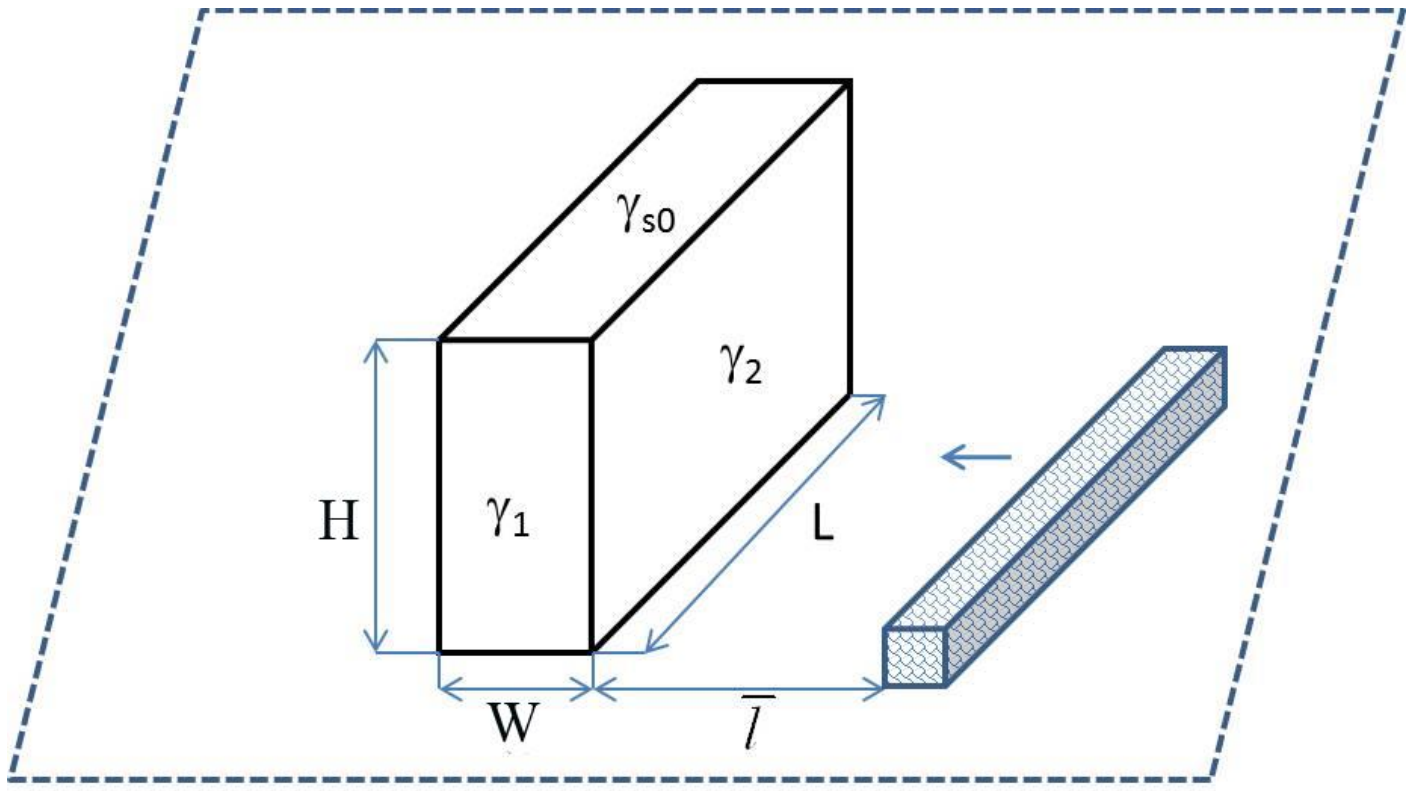

**Figure 3.** Schematic diagram of the crystal wax bar at the substrate. The relative sizes are assumed to be constant: $H/W = \kappa(\bar{\gamma}_s/\gamma_2)$, $H/L = p(\bar{\gamma}_s/\gamma_1)$

As previously discussed, thermal evaporation of thin films at room temperature introduces strain into the films. The free-energy formation of the bar can be expressed as:

$$\Delta G = (\Delta G_v + \sigma_\varepsilon)\cdot WHL + 2(WH)\gamma_1 + 2(HL)\gamma_2 + WL(\gamma_s + \gamma_{s0}) \quad (2)$$

where $\Delta G_v$ is the volumetric free-energy formation of $C_{36}H_{74}$, and $\sigma_\varepsilon$ is the deformation energy per unit volume of the strained bar.

Using the bar's size ratios:

$$\Delta G = \left(\Delta G_v + \frac{2\overline{\gamma}}{\widehat{R}} + \sigma_\varepsilon\right)\left(HWL\right) \qquad (3)$$

where $\widehat{R} = \left(V_0 \overline{\gamma}_s^2 / \gamma_1 \gamma_2\right)^{1/3}$ and $\overline{\gamma} = \overline{\gamma}_s (1+\kappa+p)/\left(\kappa p\right)^{1/3}$ are respectively the effective size and the effective surface energy of the bar, $H = \widehat{R}\left(\kappa p\right)^{1/3}$. The wax bar can be described as a conformation of N long molecular chains of length *l* and unit volume $v_1$; thus, $Nv_1 = WHL = \widehat{R}^3 (\gamma_1 \gamma_2 / \overline{\gamma}_s^2)$. The average chemical potential of a single molecular chain inside the bar can be written as:

$$\mu(\widehat{R},\varepsilon) = \frac{\partial \Delta G}{\partial N} = \mu_{0w} + \frac{4\overline{\gamma} v_1}{3\widehat{R}} + \sigma_\varepsilon v_1 \qquad (4)$$

where $\mu_{0w}$ is the chemical potential of single molecular chains in the large wax crystal. The chemical potential of the single molecular chains on the substrate immediately adjacent to the bar's surface, $\mu_s$, is equal to the chemical potential inside the bar, $\mu(\widehat{R},\varepsilon)$:

$$\mu_s = \mu_{0s} + kT \ln C_{\widehat{R}} = \mu(\widehat{R},\varepsilon) = \mu_{0w} + \frac{4\overline{\gamma} v_1}{3\widehat{R}} + \sigma_\varepsilon v_1 \qquad (5)$$

where $\mu_{0s}$ is the chemical potential of a single molecular chain on the substrate far from the wax crystals. From eq. (5) we can find the equilibrium concentration of single molecular chains at the substrate surface adjacent to a bar of effective size $\widehat{R}$:

$$C_{\widehat{R}} = C_{we} \exp\left(\frac{4\overline{\gamma} v_1}{3\widehat{R}kT} + \frac{\sigma_\varepsilon v_1}{kT}\right) \equiv B_\varepsilon C_{we} \exp\left(\frac{4\overline{\gamma} v_1}{3\widehat{R}kT}\right) \qquad (6)$$

where $C_{we} = \exp\left(\frac{\mu_{0w} - \mu_{0s}}{kT}\right)$ is the equilibrium concentration near the large wax crystals ($\widehat{R} \to \infty$) and $\mathrm{B}_\varepsilon \equiv \exp\left(\frac{\sigma_\varepsilon v_1}{kT}\right)$.

The growth rate of the bar is proportional to the flux of singular chain-like $C_{36}H_{74}$ molecules along the surface to the long and short sides of the bar:

$$\frac{dV_{bar}}{dt} = 3\widehat{R}^2\left(\frac{\gamma_1\gamma_2}{\overline{\gamma}_s^2}\right)\frac{d\widehat{R}}{dt} = -2J_{bar}\cdot v_1\left(L+W\right)\lambda = -2J_{bar}\cdot\widehat{R}\cdot(\kappa p)^{1/3}\left(\frac{\gamma_1}{p\overline{\gamma}_s}+\frac{\gamma_2}{\kappa\overline{\gamma}_s}\right)\lambda v_1 \qquad (7)$$

which is reduced to the expression:

$$\widehat{R}\frac{d\widehat{R}}{dt} = -J_{bar}v_1\lambda f(\kappa, p) \qquad (7a)$$

where $f(\kappa,p) = (2/3)(\kappa p)^{1/3}\overline{\gamma}_s\left(1/\gamma_2 p + 1/\gamma_1\kappa\right)$, and $\lambda$ is the thickness of a diffusion layer (Figure 4). The flux $J_{bar}$ is the number of molecules of the volume $v_1$ that pass through the unit interface of the bar in 1 s. The flux is determined by diffusion of single molecular chains along the surface of the substrate and absorption of these chains by the wax bar at the bar's interface. The latter process may be related to overcoming a substantial barrier known as the Ehrlich-Schwöebel barrier (ESB). In a steady-state process, the diffusion flux along the substrate's surface, $J_D$, should be equal to the flux through the ESB, $J_{ESB}$ (Figure 4):

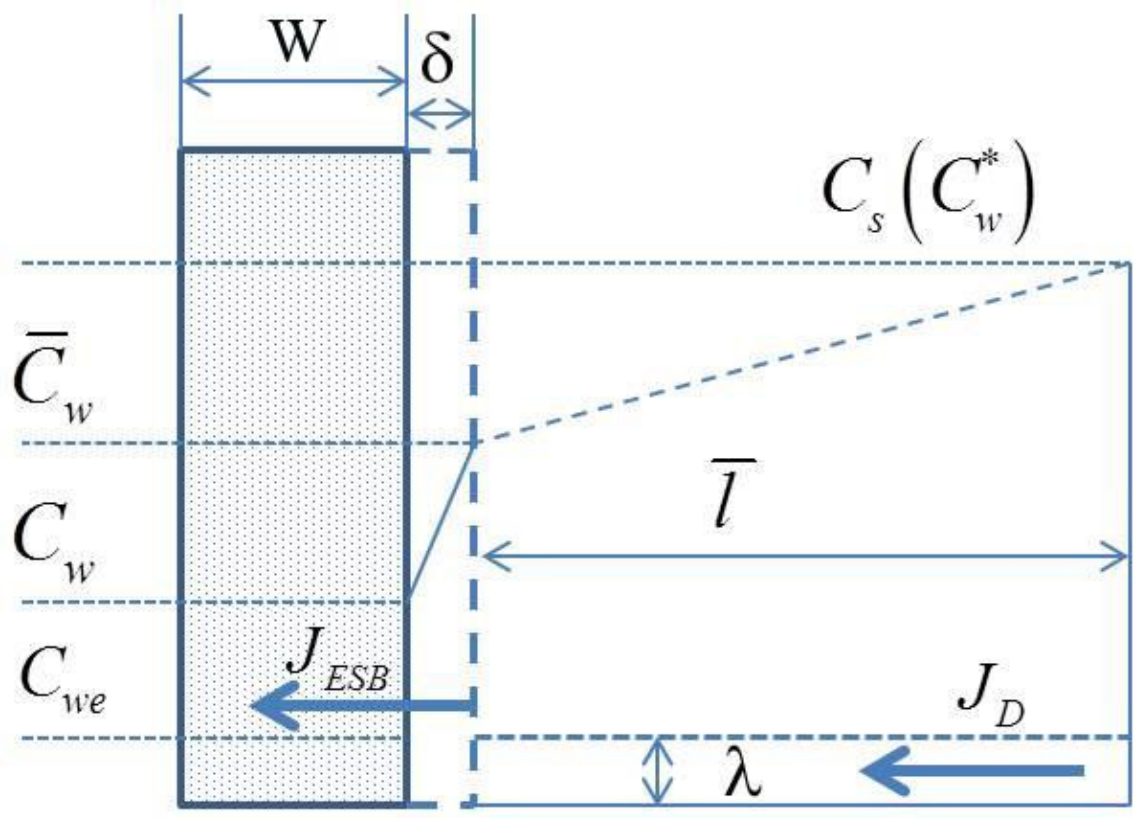

**Figure 4.** Schematic diagram of the concentration profile of long-chain alkanes at the substrate surface near the crystal wax bar

$$v_1 J_{bar} = v_1 J_{ESB} = -D_b \frac{\left(\bar{C}_{\hat{R}} - C_{\hat{R}}\right)}{\delta} = v_1 J_D = D_s \frac{\left(C_s - \bar{C}_{\hat{R}}\right)}{\bar{l}} = \frac{D_s}{\bar{l}} \frac{\left(C_s - C_{\hat{R}}\right)}{(1+\beta)} \tag{8}$$

where $D_s$ is the diffusivity of single-chain molecules at the substrate surface, $D_b$ is the diffusion coefficient through ESB, $\bar{l}$ is the characteristic distance near the bar where the concentration changes from $C_{\hat{R}}$ to $C_s$, $\bar{C}_{\hat{R}}$ is the stationary concentration at the wax crystal interface with the effective size $\hat{R}$, $C_s$ is the concentration at the surface far from the bars, and δ is the thickness of the wax interface layer, $\beta = \frac{D_s}{D_b}\frac{\delta}{\bar{l}}$. Assuming cylindrical symmetry near the bar, the stationary concentration profile can be found from the Laplacian equation:

$$\nabla^2 C = \frac{1}{r}\frac{\partial}{\partial r}\left(r\frac{\partial C}{\partial r}\right) = 0 \tag{9}$$

with the boundary conditions:

$$\begin{aligned} &C(r=\hat{R}) = \bar{C}_{\hat{R}}; \\ &C(r=\bar{l}) = C_s \end{aligned} \tag{10}$$

This results in the concentration distribution:

$$C = C_{\hat{R}} + \frac{C_s - \bar{C}_{\hat{R}}}{ln(\bar{l} \,/\, \hat{R})} ln(r \,/\, \hat{R}) \tag{11}$$

The flux $J_{bar}$ is determined by the expression:

$$v_1 J_{bar} == v_1 J_D = D_s \left.\frac{\partial C}{\partial r}\right|_{r=\hat{R}} = \frac{D_s}{\hat{R}} \frac{\left(C_s - \bar{C}_{\hat{R}}\right)}{\ln(l_b / \hat{R})} = \frac{D_s}{(1+\beta)\hat{R}} \frac{\left(C_s - C_{\hat{R}}\right)}{\ln(l_b / \hat{R})} \tag{12}$$

where $\left(2l_b\right)$ is the average distance between the bars. For $C_s >> C_{\hat{R}}$, evolution of the bar's effective size is described by the following equation:

$$Q\hat{R}^2 \frac{d\hat{R}}{dt} = \frac{2\lambda D_s C_s}{1+\beta} f(\kappa, p) \tag{13}$$

where $\mathrm{Q} = \ln(l_b / \hat{R})$. Assuming the average aspect ratios of the bars to be constant, and neglecting the variation of the logarithmic function, $Q \approx const$, a cubic-like growth law can be obtained:

$$\hat{R}^3(t) - \hat{R}^3(0) = \frac{6\lambda D_s C_s}{Q\left(1+\beta\right)} f(\kappa, p)t \tag{14}$$

At the later stage (coarsening) the concentration $C_s = C^*_{\hat{R}}$ far from the bars corresponds to bars of the critical size $\hat{R}^*$, which separates growing from dissolving bars. In this case, the flux can be evaluated as:

$$J_{bar} = \frac{D_s}{\left(1+\beta\right)} \frac{C_{\hat{R}} - C^*_{\hat{R}}}{Qv_1\hat{R}} = \frac{D_s}{\left(1+\beta\right)} \frac{C_{we}}{Qv_1\hat{R}} \left[ B_\varepsilon \exp\left(\frac{w_0}{\hat{R}}\right) - \exp\left(\frac{w_0}{\hat{R}^*}\right) \right] \tag{15}$$

where $w_0 = \dfrac{4\bar{\gamma} v_1}{3kT}$ is a characteristic length. Evolution of the bar's size can now be described by the equation:

$$\frac{d\hat{R}}{dt} = \frac{1}{\hat{R}^2} \frac{D_s C_{we} \lambda}{\left(1+\beta\right)Q} f(\kappa, p) \left[ \exp\left(\frac{w_0}{\hat{R}^*}\right) - B_\varepsilon \exp\left(\frac{w_0}{\hat{R}}\right) \right] \tag{16}$$

Let's assume that the maximum growth rate corresponds to the average effective size of the bar, $\bar{R}$. The maximum rate can be found from the condition:

$$\left.\frac{\partial\left(d\hat{R}/dt\right)}{\partial\hat{R}}\right|_{\bar{R}}=0 \tag{17}$$

The relation between $\bar{R}$ and $\hat{R}^*$ follows from equations (16) and (17):

$$\exp\left(\frac{w_0}{\hat{R}^*}\right)=B_\varepsilon\left(1+\frac{w_0}{2\bar{R}}\right)\exp\left(\frac{w_0}{\bar{R}}\right) \tag{18}$$

By substituting (18) in (16), the equation for the evolution of the average bar size can be found:

$$\bar{R}^3\frac{d\bar{R}}{dt}=\frac{D_sC_{we}w_0\lambda}{2Q\left(1+\beta\right)}f(\kappa,p)\exp\left(\frac{w_0}{\bar{R}}\right)B_\varepsilon \tag{19}$$

For $q\equiv w_0/\bar{R}<<1$and constant values of κ and p, we can find the relation:

$$\bar{R}^4(t)-\bar{R}^4(0)=\frac{2D_sC_{we}w_0\lambda}{Q\left(1+\beta\right)}f(\kappa,p)\exp\left(\frac{\sigma_\varepsilon v_1}{kT}\right)t\equiv\left(\alpha_{PPK}t\right)B_\varepsilon \tag{20}$$

where $\alpha_{PPK}\equiv\frac{2D_sC_{we}w_0\lambda}{Q\left(1+\beta\right)}f(\kappa,p)$ can be defined as the wax crystal interface coarsening constant. The fourth degree root growth law $\bar{R}(t)\sim t^{1/4}$ is well known for interface diffusion-controlled coarsening [11].

In the general case, the average aspect ratios of the bars, and consequently the values of κ and p, vary during annealing, being a function of time and temperature. Finding these dependencies may allow exact integration of equation (19). For large dimensionless times $\tau=\alpha_{PPK}B_\varepsilon t/w_0^3>>1$ and constant κ and p, the equation returns to quaternary dependence (20) (see SI part A). For $\tau\leq 1$, however, dependence can differ substantially from 1 (Figure S1).

The value of $\alpha_{PPK}$ can be estimated from the experimental data on roughness evolution during annealing at different temperatures, and thus the diffusion coefficient and activation energy of the process can be found.

÷APPLICATION OF THE MODEL TO EXPERIMENTAL RESULTS

We measured the root mean square (RMS) nano-roughness, $H_r$, of the thermally deposited n-hexatriacontane wax films during different annealing temperatures.[18] The results can be now analyzed in accordance with the coarsening kinetics (equations 13 and 19), taking into account that the nano-roughness corresponds to the average height of the islands above the continuous wax film, which grow during coarsening and transform to the observed platelet-like wax crystals, $\bar{\mathrm{H}} = (\kappa \mathrm{p})^{1/3} \bar{\mathrm{R}}$ .

The evolution of RMS nano-roughness during annealing at different temperatures is shown in Figure 5. The slope of the functions $H_r(t)$ in double logarithmic coordinates (Figure 5a) closes to s = 1/4 after a certain transition period that decreases with temperature. The functions $H_r^4(t)$ are well approximated (Figure 5b) (much better than the functions $H_r^3(t)$) by the linear dependencies, which are indicative of the interface coarsening regime over the long enough annealing duration. The characteristic time for transition from a cubic-like growth to the quaternary-like coarsening can be estimated as $\bar{t} = w_0^3 / \alpha_{PPK} B_\varepsilon$: when annealing time $t >> \bar{t}$ , the above-introduced dimensionless time $\tau >> 1$. Based on evaluated parameters (see below) we can find $\bar{t}$ for different annealing temperatures (see Table 2). At room temperature (25 ºC) $\bar{t} \approx 4$ min, and is smaller for higher temperatures.

As demonstrated below (see SI part B), the transition behavior of the slope is related to the change in the wax crystal aspect ratios during the initial period of annealing.

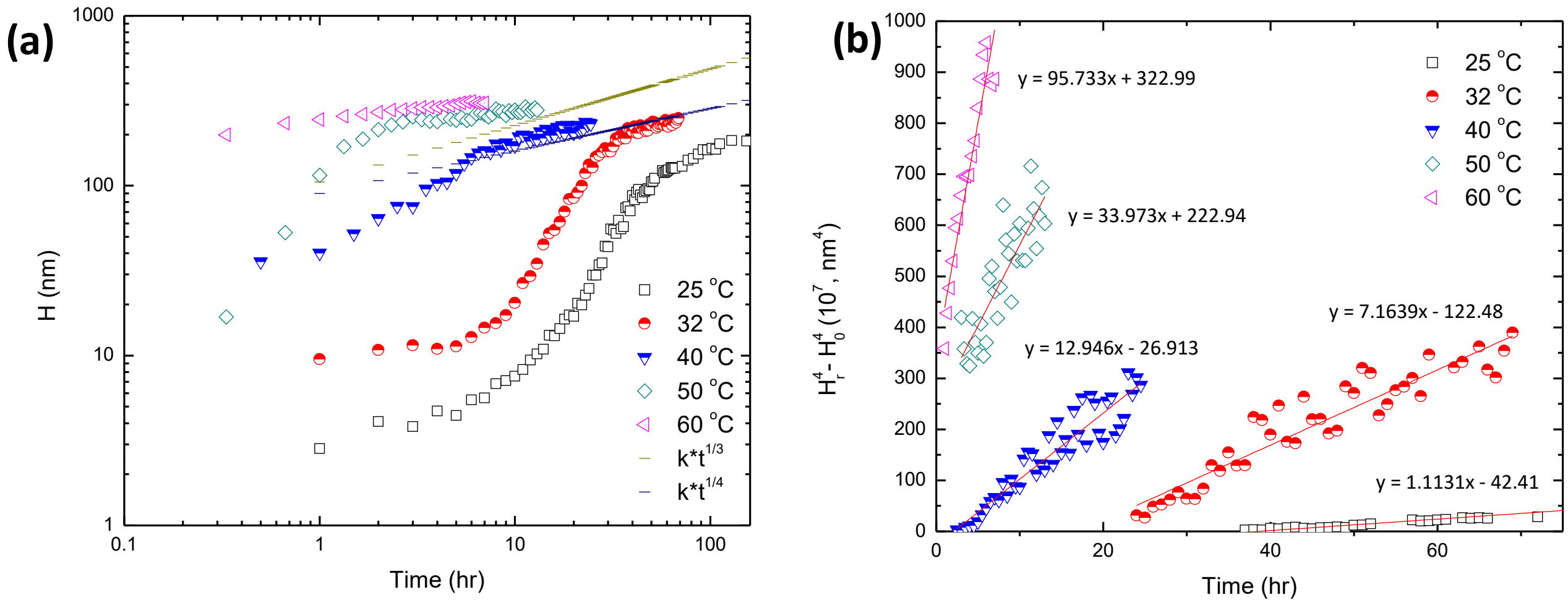

**Figure 5.** RMS nano-roughness, $H_r$, of the thermally deposited n-hexatriacontane films during annealing at (25−60) °C for different times: (a) slopes of the dependencies $H_r$ (t) approaching the value 1/4 for later times; (b) linear approximations of $\left(H_r^4 - H_0^4\right)$ for later annealing times (after transition periods).

We used the temperature dependence of the slope of linear approximations to determine the activation energy of the coarsening process (**Figure 6**), $E_{act}$ ≈1.0 eV.

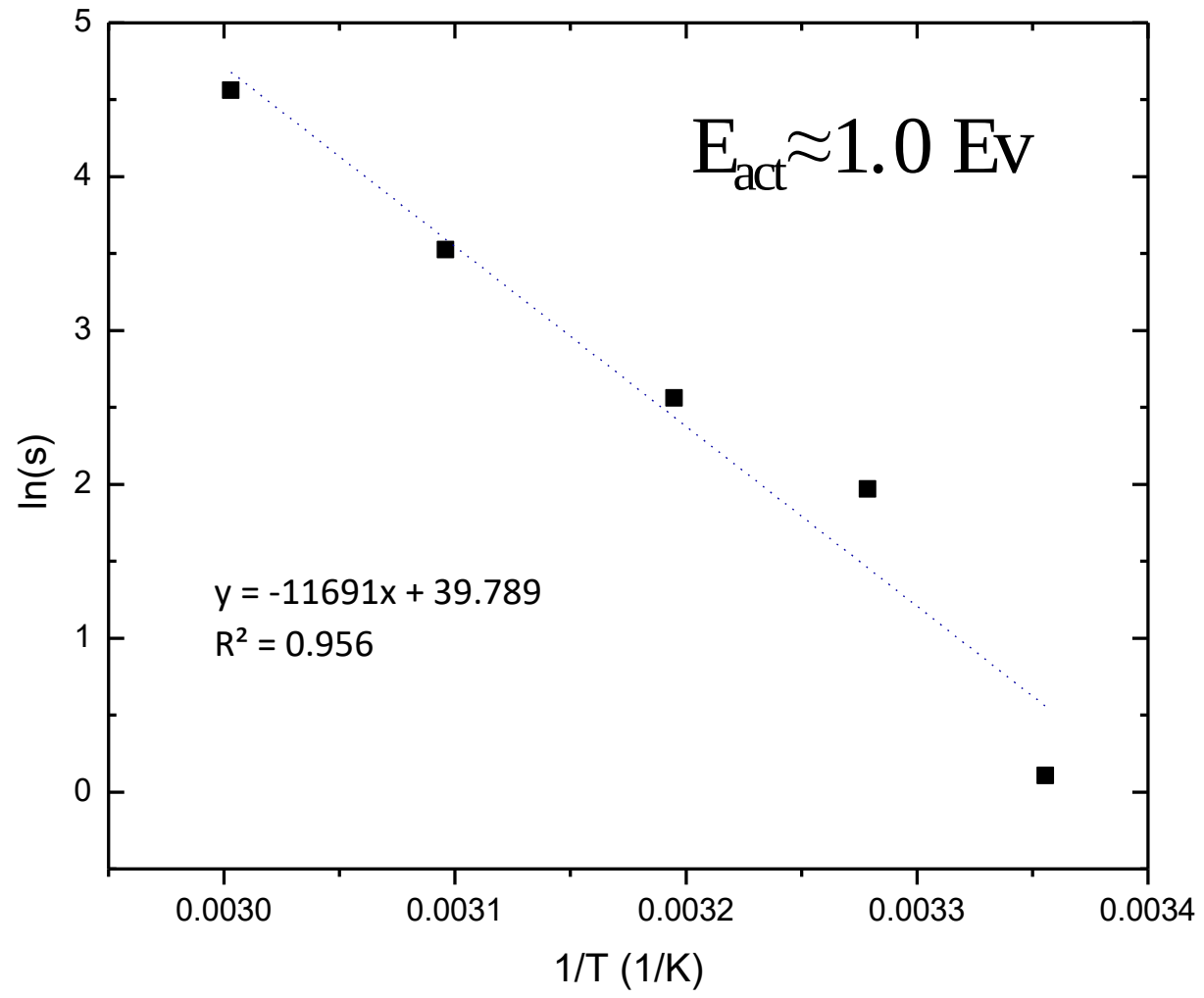

**Figure 6.** Activation energy of the slope of the dependence $H_r^4(t)$ versus 1/T during annealing after thermal deposition of wax films: $E_{\mathrm{act}} \approx 1.0$ eV.

An additional factor possibly influencing the coarsening kinetics is the change in effective surface energy of the non-equilibrium wax crystals, $\bar{\gamma} = \bar{\gamma}_s(1+\kappa+p)/(\kappa p)^{1/3}$, since the aspect ratios of the wax crystals vary during annealing, tending towards equilibrium, when $\kappa = p = 1$. By using the experimental time dependencies of aspect ratios $H/W$ and $H/L$, we were able to find the values of $(\bar{\gamma}_s/\gamma_2) \approx 4.36$ and $(\bar{\gamma}_s/\gamma_1) \approx 0.62$. The measured normalized aspect ratios as a function of time during annealing at 40 °C and 50 °C after thermal deposition are presented in Figure. 7.

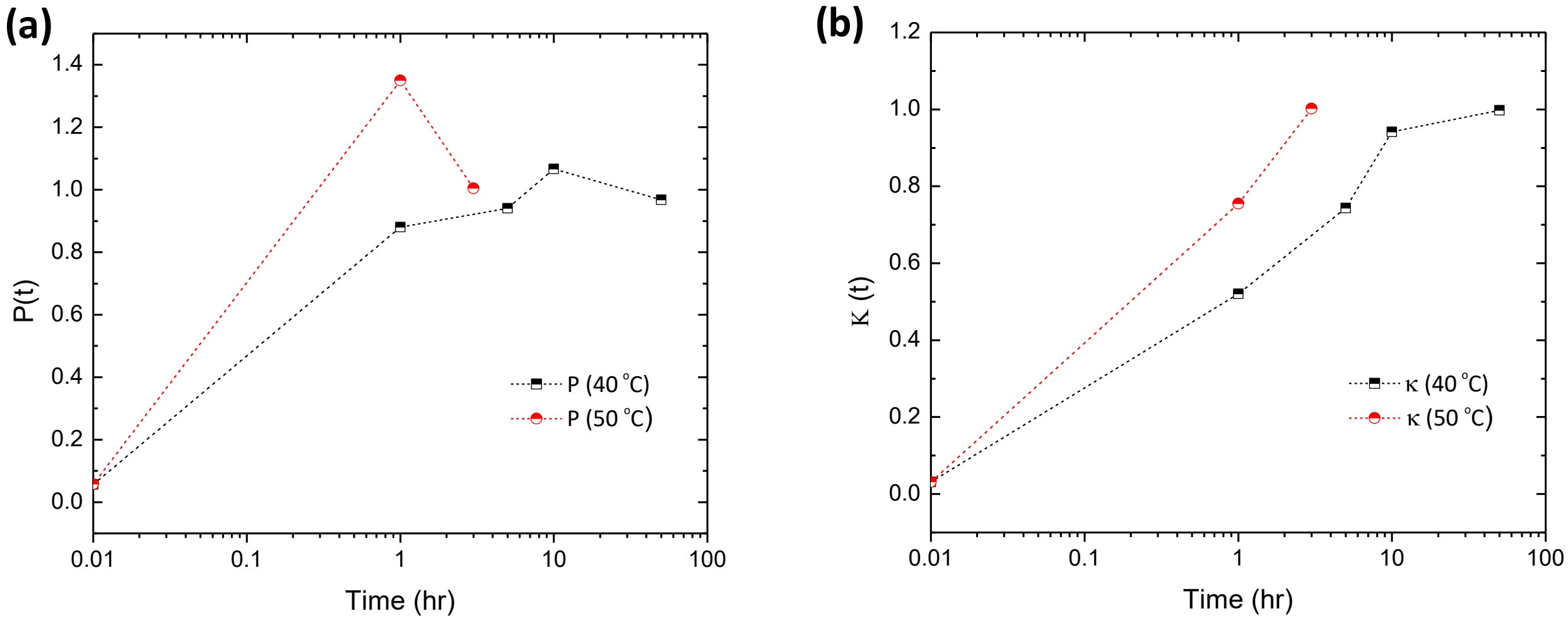

**Figure 7.** Experimental change in aspect ratios during annealing at 40 °C and 50 °C: (a) $p(t)=(H/L)/(\bar{\gamma}_s/\gamma_1)$; (b) $\kappa(t)=(H/W)/(\bar{\gamma}_s/\gamma_2)$; $(\bar{\gamma}_s/\gamma_2)=4.36$ and $(\bar{\gamma}_s/\gamma_1)=0.62$

DISCUSSION

The evolution of wax film roughness during annealing follows quaternary-like dependence of the mean average size on time, corresponding to interface diffusion-controlled coarsening. The driving force of this process is the decrease in total surface energy. The activation energy of the coarsening process of strained wax crystals can be expressed as follows:

$$E_{act}^{st} = E_{dif} + \Delta h_s - \sigma_\varepsilon v_1 \qquad (21)$$

The first term in eq. (21) is the diffusional barrier, the second is the sublimation enthalpy of n-alkanes from the wax crystal to the substrate surface, and the third is the strain energy per one molecular chain in the strained bar. Weber et al.[19] investigated the growth dynamics of n-alkanes on silica at 35 °C. While they did not find the absolute value of the diffusional barrier, they showed that the increase in the barrier per additional carbon atom is ~ 0.5 meV. This corresponds to the diffusion barrier $E_{dif}$ = 0.2 eV for n-alkane $C_{36}H_{74}$ chains.

The Ehrlich-Schwöebel barrier, evaluated for step-edge crossing of several organic molecules such as para-sexiphenyl molecules[20] and others[19] is $W_{ES}$=(0.08−0.8) eV. The empirical data of Mandelkern et al. for $E_D$ [21] can be summarized by the expression

$$E_D = \frac{5k_B T_m^2}{T_m - T_g} \tag{22}$$

where $T_g$ is the glass transition temperature. Using the corresponding values for n-alkane $C_{36}H_{74}$, $T_g$ = 164K[22] and $T_m$ = 347K, we obtain $E_D$ = 0.28eV. Therefore, the diffusional barrier and the Ehrlich-Schwöebel barrier are of the same order. In such a case, we can conclude that $D_s\delta << D_b\bar{l}$ , i.e. β << 1, and that the mass transport is controlled by diffusion along the surface of the substrate.

The second term in equation (21) is the sublimation enthalpy of n-alkanes from the wax crystal to the substrate surface. According to Bennema et al.,[23] sublimation enthalpy of the long-chain normal alkanes to vacuum for $n$ = 36 is about $\Delta h_{sv} \leq$ 35 kcal/mole =146 J/mole = 1.52 eV. To the best of our knowledge, sublimation of chain alkanes from wax crystals to a substrate has not been described. Taking into account possible binding energy of the molecule's chains to the surface, we can assume that $\Delta h_s = \theta\Delta h_{sv}$, θ < 1. Using the experimental values of $E_{act}$ ≈ 1.0 eV for activation energy and $E_{dif}$ =(0.2−0.3) eV for diffusional barrier, we can estimate θ = 0.5−0.6.

For evaluation of diffusion coefficients, it is convenient to write $\alpha_{PPK}^0 = D_{eff} C_{we} \bar{w}_0 \lambda$, where $D_{eff} \equiv \frac{2D_s}{Q(1+\beta)}$ is the effective diffusion coefficient and $\bar{w}_0 = \frac{8\bar{\gamma}_s \nu_1}{9k_B T}$. The equilibrium concentration can be evaluated by the expression:

$$C_{we} = \exp\left(\frac{\mu_{0w} - \mu_{0s}}{kT}\right) \approx \exp\left(-\frac{\Delta g_s}{RT}\right) = \exp\left[-\frac{\Delta h_s - T\Delta s_s)}{RT}\right] \tag{23}$$

where $\Delta g_s$, $\Delta h_s$ and $\Delta s_s$ are respectively the molar free energy, enthalpy and entropy of sublimation of long chains from the wax crystal to the substrate surface. The entropy of sublimation can be estimated as $\Delta s_s = \theta \Delta s_f = \theta \Delta h_f / T_m$, where $\Delta h_f$ and $\Delta s_f$ are the enthalpy and entropy of fusion and $T_m$ is the melting temperature of the wax. Using the value $\Delta h_f$ = 116 kJ/mole for n-alkanes with n = 36,[24] we can estimate $C_{we}$ for different values of θ (Table 1).

**Table 1.** Possible equilibrium concentrations of long-chain n-alkanes at the substrate $C_{we}$

| $C_{we} = \exp\left[(\theta / R)\left(\Delta h_f / \mathrm{T}_m - \Delta h_{sv} / T\right)\right], \quad \times 10^{-5}$ | | | | |
|---|---|---|---|---|
| T, K | θ = 0.5 | θ = 0.6 | θ = 0.7 | θ = 0.8 |
| 298 | 8.6 | 1.3 | 0.20 | 0.03 |
| 305 | 16.9 | 3.0 | 0.52 | 0.09 |
| 313 | 35.3 | 7.2 | 1.47 | 0.30 |
| 323 | 84.2 | 20 | 5.0 | 1.2 |
| 333 | 190 | 54 | 15.5 | 4.4 |

The values $\alpha_{PPK}^0 = S / \left(B_\varepsilon \cdot \kappa p \cdot \bar{\psi}\right)$ are proportional to the experimental values of slope *S*, corrected by the shape correction function $\bar{\psi}$ and the aspect ratios product $(\kappa p)$ (Table 2). For reasonable values: $v_1 = abl = (0.5 \cdot 0.7 \cdot 9.5) nm \simeq 3.3 \cdot 10^{-27} m^3$, $\bar{\gamma}_s = (30-50)\,\mathrm{m} J / m^2$, T=300K, $\bar{w}_0 = (30-50)\,\mathrm{nm}$, λ ≈ 1 nm and $C_{we} D_{eff} \approx (2-200)\, nm^2 / s$ (see Table 2, where evaluations were performed assuming $\bar{\gamma}_s = 50\,\mathrm{m} J / m^2$ θ = 0.5).

**Table 2**. Experimental and calculated kinetic parameters of the wax crystal films

| T, K | S $10^4$ $nm^4/s$ | $\alpha^0_{PPK}$ $10^2$ $nm^4/s$ | $B_\varepsilon$ | $C_{we}$ $\times 10^{-4}$ | $\lambda C_{we} D_{eff}$ $nm^3/s$ | $\bar{t}$, s |
|---|---|---|---|---|---|---|
| 298 | 0.31 | 0.6 | 3.34 | 0.9 | 1.6 | 235 |
| 305 | 1.99 | 4.0 | 3.25 | 1.7 | 11.0 | 34 |
| 313 | 3.60 | 7.4 | 3.15 | 3.5 | 21.0 | 17 |
| 323 | 9.44 | 17.8 | 3.05 | 8.4 | 52.3 | 6.8 |
| 333 | 26.59 | 51.9 | 2.94 | 19.1 | 157.1 | 2.2 |

Using the estimated values of $C_{we}$ (Table 2) we can evaluate the diffusion coefficient of single-chain molecules along the substrate surface as $(2-8.5)\cdot 10^{-14}$ $m^2/s$ at 25−60 ºC.

As mentioned above, two additional factors influencing the kinetics of coarsening are the strains introduced into the wax crystals during thermal deposition at room temperature and an initial non-equilibrium shape of the wax crystals. The deformation introduced during deposition was found to be about $\varepsilon \approx 1\%$.[18] These strains may be caused by the substantially different thermal expansion coefficients of paraffin wax and that of the substrate materials (for paraffin it is about $10^{-4}$ $K^{-1}$ while for Si it is $4.5\cdot 10^{-6}$ $K^{-1}$). Let us now estimate possible values of deformation energy $(\nu_1 \sigma_\varepsilon)$ for such deformations. Interatomic potentials in the n-hexatriacontane crystals are expressed by the Buckingham potential:[23]

$$\Phi(r) = \frac{A}{r^6} + Be^{-Cr} \qquad (24)$$

where r is the interatomic distance. Strain of the bar, ε, can be considered as a change of interatomic distance to a small value $\Delta r = \varepsilon r$, which results in change of the interatomic bonds:

$$\Delta\Phi = \frac{d\Phi}{dr}\Delta r = -\varepsilon\left(\frac{6A}{r^6} + BCre^{-Cr}\right) \quad (25)$$

Use of the values of constants[23] A = 535 kcal·mol$^{-1}$Å $^{6}$, B = 74460 kcal·mol$^{-1}$, C = 3.6 Å$^{-1}$, and $r = \sqrt{\left(a^2 + b^2\right)} / 4$, a = 4.97Å, b = 7.478Å, yields $\Delta\Phi \approx -0.5\varepsilon$ kcal·mol$^{-1}$. Each atom has 4 strong bonds in the lateral directions, and the overall bond energy of the molecular chain is a linear function of its length, n. The strain energy caused by deformation ε can be evaluated as

$$\sigma_\varepsilon v_1 = 4n\left|\Delta\Phi\right| \approx 2n\varepsilon \text{ kcal/mol} \quad (26)$$

For n = 36 and ε = 0.01, we find $\sigma_\varepsilon v_1$ = 0.72 kcal/mol = 0.031 eV. This corresponds to a decrease in effective activation energy of the coarsening process to a value ~ 0.031 eV. The factor $B_\varepsilon$ for T = (298K−323K) where $B_\varepsilon = \exp\left(\sigma_\varepsilon v / k_B T\right) = \left(3.3 - 2.9\right)$, means that the rate of coarsening can be ~3 times faster in strained than in unstrained n-hexatriacontane crystals. During annealing the decreases in strain might result in a corresponding slight increase in activation energy.

The second substantial factor influencing the coarsening kinetics of thermally deposited films is an increased effective surface energy of the non-equilibrium wax crystals. During very slow deposition the shape of wax crystals is dictated by minimization of the surface energy, while fast deposition does not allow such minimization. After fast deposition the wax crystals are usually shorter and wider than wax crystals at equilibrium. During annealing the crystal shape evolves towards the shape at equilibrium. This process is controlled by diffusion of singular chain-like $C_{36}H_{74}$ molecules along the surfaces of the wax bars, while at the same time this evolution competes with the change in shape due to the arrival of new molecules from the substrate surface. The result is a non-monotonic change of aspect ratios during annealing (Figure 7).

For wax crystals that are initially wider ($\kappa < 1$) and shorter ($p > 1$) than those at equilibrium, the rate of evolution during thermal annealing can be significantly faster than in equilibrium shaped crystals, since the effective surface energy $\bar{\gamma} = \bar{\gamma}_s (1+\kappa+p)/(\kappa p)^{1/3}$ becomes substantially higher than the equilibrium value0.07 = κand 6p = 1. . For example, for $\bar{\gamma}_{eq} = 3\bar{\gamma}_s$ (typical dimension ratios for initial wax crystal shape after fast deposition) the effective surface energy $\bar{\gamma} = 5.54\bar{\gamma}_s$. The value $w_0 \sim \bar{\gamma}$ and the shape function $f(p,\kappa)$ (in equations (14), (19)) become temperature- and time-dependent during the annealing process since the aspect ratios tend to be equilibrium values ($\kappa$, $p \rightarrow 1$). This may affect the initial kinetics of the bar's growth, especially at lower annealing temperatures (25−32 ºC). The variation in slope during the transition period of annealing may be related to the change in aspect ratios of the wax crystals during this period (Figure S3, SI part B).

The kinetics of shape equilibration of the wax is dictated by its surface mass flows, which are driven by the difference in mean chemical potentials at the top and back facets of the bar: $\mu_1 = 2\nu_1\gamma_1 / L, \mu_2 = 2\nu_1\gamma_2 / w$ and $\mu_s = 2\nu_1\bar{\gamma}_s / h$, defined in terms of weighted mean curvature.[25] The total mass arrival rate for a facet is the sum of the mass flow rates at the facet edges. These flow rates are dictated by the chemical potential gradients at the facet edges. When facets undergo uniform normal displacements that preserve their shape and orientation, the gradient or divergence of the flux must be constant on each facet, resulting in parabolic variation of the chemical potential along the facets.[25,26] The exact solution of this problem is beyond the scope of the present investigation. At the same time, our experimental investigation of the changes in wax crystal shape during annealing at different temperatures allowed us to ascertain and approximate the time and temperature dependencies of the ratios p and κ, and use them to evaluate the wax coarsening process.

## CONCLUSIONS

Hexatriacontane ($C_{36}H_{74}$) wax crystal films that are thermally deposited on a silicon substrate undergo a roughening process during annealing above room temperatures. Coarsening of the $C_{36}H_{74}$ crystal islands responsible for the measured nano-roughness of the wax films, is driven by the decrease in total surface energy. The coarsening kinetics are controlled by diffusion of single molecular chains along the substrate that provides a linear time dependence of the forth degree of the nano-roughness, $H_r^4(t) \sim t$. Two additional factors influence the evolution rate: strains accumulated in the wax crystals during deposition and the initial shape of the crystals (divergence from the shape at equilibrium). Both factors lower activation energy and effectively shorten the coarsening time. Variations in the aspect ratios of the wax crystals lead to a decrease in the slope of the dependence $H_r^4(t)$ during the transition period of annealing. Our model of paraffin wax crystal coarsening provides a good description of the experimental results and makes it possible to predict the behavior of paraffin films based on the extracted kinetic parameters of the system.

## ASSOCIATED CONTENT

**Supporting Information**

Modeling of the initial stages of coarsening (SI Part A) and description of the influence of the bar's shape change during annealing (SI Part B). This material is available free of charge via the Internet at http://pubs.acs.org.

AUTHOR INFORMATION

**Corresponding Author**

*Email: bpokroy@tx.technion.ac.il, akatsman@tx.technion.ac.il

**Author Contributions**

The manuscript was written through contributions of all authors. All authors have given approval to the final version of the manuscript.

ACKNOWLEDGMENT

The research leading to these results received partial funding from the European Research Council under the European Union's Seventh Framework Program (FP/2007–2013)/ERC Grant Agreement no. 336077) and partial funding by Grant 421-0251-13 from the Office of the Chief Scientist of the Ministry of Agriculture (Israel).

The authors thank Lihi Shenhav and Dennis Markovic for their contribution in roughness measurements.

REFERENCES

(1) Zasadzinski, J.; Viswanathan, R.; Madsen, L.; Garnaes, J.; Schwartz, D. *Science* **1994**, *263*, 1726.
(2) Kind, M.; Wöll, C. *Prog. Surf. Sci.* **2009**, *84*, 230.
(3) Kraus, M.; Haug, S.; Brütting, W.; Opitz, A. *Org. Electron.* **2011**, *12*, 731.

(4) Kraus, M.; Richler, S.; Opitz, A.; Brütting, W.; Haas, S.; Hasegawa, T.; Hinderhofer, A.; Schreiber, F. *J. Appl. Phys.* **2010**, *107*, 094503.
(5) Luff, P.; White, M. *Thin Solid Films* **1970**, *6*, 175.
(6) Fu, J.; Urquhart, S. G. *Langmuir* **2007**, *23*, 2615.
(7) Tanaka, K.; Okui, N.; Sakai, T. *Thin Solid Films* **1991**, *196*, 137.
(8) Ashida, M.; Ueda, Y.; Yanagi, H. *Bull. Chem. Soc. Jpn* **1986**, *59*, 1437.
(9) Forrest, S. R. *Chem. Rev.* **1997**, *97*, 1793.
(10) Beernink, G.; Strunskus, T.; Witte, G.; Wöll, C. *Appl.Phys. Lett.* **2004**, *85*, 398.
(11) Kowarik, S.; Gerlach, A.; Schreiber, F. *J. Phys. Condens. Matter* **2008**, *20*, 184005.
(12) Yim, S.; Jones, T. *Phys. Rev. B* **2006**, *73*, 161305.
(13) Magonov, S. N.; Yerina, N. A. *Langmuir* **2003**, *19*, 500.
(14) Shimizu, H.; Tanigaki, N.; Nakayama, K. *Jpn. J. Appl. Phys.* **1995**, *34*, L701.
(15) Ishida, K.; Hayashi, K.; Yoshida, Y.; Horiuchi, T.; Matsushige, K. *J. Appl. Phys.* **1993**, *73*, 7338.
(16) Craig, S. R.; Hastie, G. P.; Roberts, K. J.; Sherwood, J. N. *J. Mater. Chem.* **1994**, *4*, 977.
(17) Leunissen, M. E.; Graswinckel, W. S.; van Enckevort, W. J. P.; Vlieg, E. *Cryst. Growth Des.* **2004**, *4*, 361.
(18) Pechook, S.; Pokroy, B. *Adv. Funct. Mater.* **2012**, *22*, 745.
(19) Weber, C.; Frank, C.; Bommel, S.; Rukat, T.; Leitenberger, W.; Schäfer, P.; Schreiber, F.; Kowarik, S. *J. Chem. Phys.* **2012**, *136*, 204709.
(20) Hlawacek, G.; Puschnig, P.; Frank, P.; Winkler, A.; Ambrosch-Draxl, C.; Teichert, C. *Science* **2008**, *321*, 108.
(21) Mandelkern, L. *Polym. Eng.Sci.* **1969**, *9*, 255.
(22) Waheed, N.; Ko, M.; Rutledge, G. *Polymer* **2005**, *46*, 8689.
(23) Bennema, P.; Liu, X. Y.; Lewtas, K.; Tack, R.; Rijpkema, J.; Roberts, K. *J. Cryst. Growth* **1992**, *121*, 679.
(24) Lira-Galeana, C.; Firoozabadi, A.; Prausnitz, J. M. *AIChE J.* **1996**, *42*, 239.
(25) Kitayama, M.; Narushima, T.; Glaeser, A. M. *J. Am. Ceram. Soc.* **2000**, *83*, 2572.
(26) Klinger, L.; Rabkin, E. *Int. J. Mater. Res.* **2010**, *101*, 75.

*For Table of Contents Use Only*

# Paraffin Wax Crystal Coarsening: Effects of Strains and Wax Crystal Shape

*Sasha Pechook, Alex Katsman, and Boaz Pokroy*

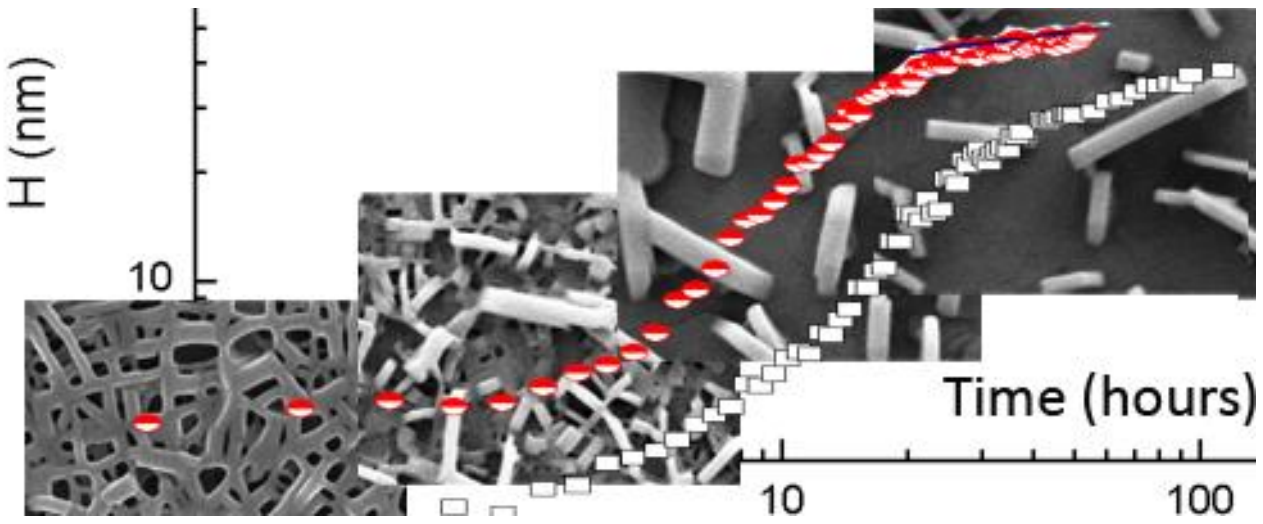

**Synopsis:** A model of paraffin wax crystal coarsening effectively describing experimental results and allowing the prediction of paraffin films behavior on the basis of the extracted kinetic parameters was developed. Evolution rate was influenced by strains accumulated in wax crystals during deposition and divergence of initial crystal shape from equilibrium shape. Both factors lowered activation energy and effectively shortened coarsening time.